\documentclass{IEEEcsmag}
\usepackage[T1]{fontenc}
\usepackage{helvet}
\usepackage[colorlinks,urlcolor=blue,linkcolor=blue,citecolor=blue]{hyperref}
\expandafter\def\expandafter\UrlBreaks\expandafter{\UrlBreaks\do\/\do\*\do\-\do\~\do\'\do\"\do\-}
\usepackage{upmath,color}

\jvol{XX}
\jnum{XX}
\paper{8}
\jmonth{Month}
\jname{IEEE Software}
\jtitle{Generative AI in Evidence-Based Software Engineering: A White Paper}
\pubyear{2024}

\AtBeginDocument{%
  \providecommand\BibTeX{{%
    \normalfont B\kern-0.5em{\scshape i\kern-0.25em b}\kern-0.8em\TeX}}}
\usepackage{booktabs}
\usepackage[backend=biber,style=ieee,natbib=true]{biblatex}

\usepackage{graphicx}
\usepackage{multirow}
\usepackage{float}

\begin{document}


\title{Generative AI in Evidence-Based Software Engineering: A White Paper}

\author{Matteo Esposito}
\affil{University of Oulu, Oulu, Finland and University of Rome Tor Vergata, Rome, Italy}

\author{Andrea Janes}
\affil{Free University of Bozen-Bolzano, Bozen-Bolzano, Italy}

\author{Davide Taibi}
\affil{University of Oulu, Oulu, Finland}

\author{Valentina Lenarduzzi}
\affil{University of Oulu, Oulu, Finland}

\begin{abstract}
\textbf{Context}. In less than a year, practitioners and researchers witnessed a rapid and wide implementation of Generative Artificial Intelligence. The daily availability of new models proposed by practitioners and researchers has enabled quick adoption. Textual-GAI's capabilities enable researchers worldwide to explore new generative scenarios, simplifying and hastening all time-consuming text generation and analysis tasks. 
\textbf{Motivation}. The exponentially growing number of publications in our field, with the increased accessibility to information due to digital libraries, makes conducting systematic literature reviews and mapping studies an effort and time-insensitive task. Stemmed from this challenge, we investigated and envisioned the role of GAIs in evidence-based software engineering. 
\textbf{Future Directions}. Based on our current investigation, we will follow up the vision with the creation and empirical validation of a comprehensive suite of models to effectively support EBSE researchers.
\end{abstract}

\maketitle
\section{Introduction}
\label{sec:intro}
Systematic studies (SS), i.e., literature review and mapping studies,  are essential for collecting, organizing, and sharing knowledge within the research community \cite{DBLP:conf/icse/KitchenhamDJ04}. Such reviews analyze multiple studies to guide informed research and practitioner-oriented decision-making in different fields. While digital libraries and online tools offer some automation to assist researchers in finding, organizing, and grabbing studies quickly, existing tools do not fully support the entire SLR process, especially the automation of study collection, application of criteria, and data synthesis  \cite{felizardo2020automating}.

Natural language processing (NLP) research offers a state-of-the-art model that can quickly analyze the amount of text documents and pull knowledgeable information. Such a current trend materialized the concept of an AI that can create or generate, i.e., generative AI (GAI), among the possibilities, ``new'' text, art, and music based on the vast knowledge stored in the transformer network as parameters. 

Therefore, in response to the growing demand for improved research support \cite{felizardo2020automating}, and the tedious task of SS, we envision the role of GAI in evidence-based software engineering (EBSE). 
Our paper presents a vision of the role that GAI will eventually have in EBSE. Our goal is to inspire interest in integrating generative AI with systematic studies in evidence-based software engineering rather than attempting to present a comprehensive survey.
Practitioners would be interested in the evolution of the framework that will allow our vision to come to fruition and the applicability of the presented GAI approaches to industrial use cases kin to SS; researchers, on the other hand, will be interested in the performance of our suite of models and the actual speedup. More specifically, our vision describes the current use of AI in SS and combines established systematic review methodologies with current technology, such as GAI, to automate the entire review process. Our contributions are three-folded: (i) We provide a collection of studies on the current state-of-the-art leveraging GAIs,e.g., LLMs, in specific SS steps; (ii) We highlight GAI usage gaps in the current research landscape and envision methods to fill those gaps; (iii) We provide a sample empirical evaluation protocol for a comprehensive suite of GAI models for EBSE.


\section{Background \& Motivation}
\label{sec:motivation}
\citet{DBLP:conf/icse/KitchenhamDJ04} were the first to introduce an evidence-based approach in SE, and the guidelines they gave are these days the \textit{de-facto } standard.
\citet{DBLP:conf/esem/SantosMS14} investigated the benefits the SE community can gain from this approach, like in medicine. They tried to understand the evolution of systematic reviews in EBSE. \citet{DBLP:conf/esem/SantosMS14} conducted an SMS in which they discovered 52 papers that reported on the importance of SLR in software engineering research.
\citet{felizardo2020automating} study evidence-based software engineering (EBSE) by focusing on the role of SLRs. Although SLRs are at the core of EBSE, they are highly work-intensive. For the last ten years, researchers have tried to automate the process of SLRs in software engineering (SE) to decrease human effort and lessen the quality of SLRs. 
Finally, \citet{felizardo2020automating} concludes that the SE community still needs to advance and develop more automation strategies to reduce human effort further and make the SLR process smoother.

\textbf{Automated Tools}. Currently, tools that include AI functionalities for part of the SLR process, such as Data Extraction and Synthesis exists\footnote{\url{https://elicit.com}}\footnote{\url{https://www.distillersr.com/}}. Nonetheless, the current tools are proprietary. Contrary, our vision,  is not directly related to study nor focused on a systematic review, rather on presenting a structured overview of the possibilities, though some non trivial, of GAI in EBSE.

Our work proposes a vision of the role of GAIs in trying to cover all SLR activities, such as data extraction and synthesis. Hence, our approach uses a unique set of methodologies and objectives for a much broader scope in the case of SLR activity.

\section{Design}
\label{sec:design}
To our knowledge, our work classifies the latest literature on using specific GAI models for specific SS tasks. We organized the SS tasks according to the state-of-the-art guidelines \cite{keele2007guidelines}. We envision GAI to aid researchers in the following five main SS tasks. We also focus on the topic of multivocal literature review (MLR), aligning with the growing community interest, and referring to the guidelines by \citet{GAROUSI2019101}.

\textbf{Section Structure}. The following sections and subsections adhere to the guidelines provided by \citet{keele2007guidelines} as our structure for further discussion. We discuss classical methodologies and how GAI supplements or automates these methodologies.
\subsection{Research Identification}
\subsubsection{Create a search strategy}

The first step of an SLR plays a crucial role in identifying relevant studies related to the research topic or question under consideration. Researchers develop the search string using specific strategies, including predefined keywords and boolean operators, to look for relevant studies  \cite{keele2007guidelines}.

\textit{Classical Approach.} One of the classical approaches to structure a search string is to use PICO \cite{esposito2024validate,PICO}. 

PICO~\cite{PICO} is a framework borrowed from evidence-based practice, originating from healthcare, to build a search string. PICO stands for: \textbf{P}: Patient/Population/Problem; \textbf{I}: Intervention; \textbf{C}: Comparison; \textbf{O}: Outcome.

\textit{GAI Approach.} According to recent research works \cite{ghosh2024alpapico}, we suggest chat-based LLM to support PICO string creation for software engineering studies as below. Text-oriented LLMs can generate structured text following human specification \cite{esposito2024leveraging}. Therefore, they can generate PICO terms and recommend the search string of online libraries \cite{ghosh2024alpapico}. The researchers can leverage this recent GAI implementation to make canvas creation possible. So, we suggest using vision-enabled LLM to depict the search string, such as a tree, to make it easy for researchers to comprehend and modify the query accordingly. Furthermore, GAIs can search automatically for papers or articles related to the user query over several web-based libraries and combine them. This will save the researcher effort and time.




\subsubsection{Publication Bias}
Publication bias is a peculiar type commonly associated with academic research publications, where published research findings skew systematically toward positive or statistically significant ones. Simultaneously, studies with negative or non-significant findings are less likely to be published, or even if published, they are published in a less visible way \cite{keele2007guidelines}.
\textit{Classical Approach. Multivocal Litterature Review.} The MLR approach prevents publication bias by bringing various perspectives, sources, and voices into the systematic review process. It is appreciated that when the systematic review heavily relies on the published academic literature, the conclusions could be heavily biased since publication bias generally results in skewness towards positive or statistically significant results. On the other hand, MLR aims to incorporate sources and perspectives of all types, whether published or unpublished studies and web sources, to get a more profound and objective knowledge of the issue under study (a.k.a. gray literature).

\textit{GAI Approach.} LLMs, like Google's Gemini, Perplexity, or OpenAI GPT4o, are capable of surfing the web to get fresh data relevant to the user's questions. Searching for grey literature also requires many searches on search engines like Google and Yahoo, social media like YouTube and Yelp, or focused forums such as StackOverflow. 

On the other hand, gray literature is painfully non-tabulated and needs lots of formatting. \textit{Prompt engineering}, i.e., the design of the specific LLMs' prompt, allows the researchers to tell the model how to orient and make sense of the sheer amount of data, what to look for, and how to shape the output in the easiest possible way.
To the best of our knowledge, no prior study has suggested the adoption of LLM in the conduct of an SLR. Therefore, we aim to advance the state-of-the-art, adding to the empirical validation of our vision a separate section surveying the grey literature to evaluate the LLM's performance.

\subsubsection{Conducting the search}
With the research string at hand, researchers must, without exception, adjust the manually crafted search string for every digital library. These may be ACM Digital Library (ACM DL), IEEEXplore, SpringerLink, ScienceDirect, and Scopus.
\textit{Classical Approach} \textbf{}
Searching through the different digital libraries is a frustrating task. ACMDL, IEEEXplore, and Scopus have query languages, which are primarily proprietary and have the same - or very similar - field names. However, the implementation and field availability differ widely. More specifically, the three major libraries do not share the usage of, e.g., the wildcard characters '*' for matching part of the word. Moreover, not all allow copy-pasting of a query string externally crafted. 

\textit{GAI Approach} In this step, we suggest two different GAI-based approaches, given the availability of internet-enabled GAI (iGAI) \cite{sharma2024generative}. When using an \textbf{iGAI}, we see the role of, e.g., an LLM model for the role of search query creation. We suggest fine-tuning an LLM to create a chatbot for each digital library. We may fine-tune by sending pairs of messages such as search strings as user input, a search query for each digital library, and an LLM reply. \textbf{Otherwise}, an iGAI can create the search query and search by compiling and executing a predetermined script that allows API access to the digital libraries of interest. It is worth noticing that not all digital libraries allow for API access or computational access for library scraping. We address this acknowledged issue in Section \ref{sec:limitations}.  At the end of this step, researchers have the collection of primary studies collected from all the digital libraries of interest.

\subsubsection{Documenting the Search}
 According to \citet{keele2007guidelines}, conducting an SS should be transparent and replicable. Therefore, different tools and conceptual frameworks assist researchers in documenting the search protocol and findings. For instance, PRISMA, which stands for Preferred Reporting Items for Systematic Reviews and Meta-Analyses, is a set of guidelines (checklists) designed to help researchers improve the reporting of systematic reviews and meta-analyses.

\textit{Classical Approach.} 
The classical approach of documenting the search process is quite labor-intensive, in which a PRISMA framework is prepared manually in textual and graphical form. Researchers manually document each step of the search process, including the selection of digital libraries, search strings, inclusion/exclusion criteria, dates of the search, and the results of the searches. 

It is a manual process and is considerably time-consuming with a high potential for errors, as it deals with many digital libraries and iterations of search.

\textit{GAI Approach.}
GAI approach enables the automation of the PRISMA workflow. More specifically, researchers can fine-tune an LLM designed according to the PRISMA guidelines to generate the PRISMA workflow automatically. 

\subsection{Study Selection}
According to \citet{keele2007guidelines}, the study selection step identifies and includes all relevant studies that meet predefined inclusion criteria, guaranteeing that the studies being synthesized are pertinent to the researchers' investigation. Researchers establish inclusion and exclusion criteria by considering previous SS in the specific topic and based on the researchers' own research experience. According to \citet{keele2007guidelines}, usual criteria include study language, publication venue, types of participants, and research design choice.

\textit{Classical Approach.} Researchers manually review each paper examining the abstract and keywords or giving a quick read of the entire work. Based on the researchers' understandings, they exclude or include the studies for further screening.
Usually, SS encompass analyzing papers in the order of the hundreds; therefore, it is evident that the preliminary screening time can be computed at least in weeks, if not months. Moreover, more than one author must invest the same amount of time reviewing the same article for eventual disagreement. The extent of agreement among authors applying the inclusion/exclusion criteria should be computed with the usual inter-rater agreement (IRA) statistic such as Kendal's $\tau$ or Cohen's $\kappa$ .

\textit{GAI Approach.} According to \citet{10.1145/3613904.3642216}, LLM can be fine-tuned to apply user-defined criteria to prompts. Nonetheless, to our knowledge, no prior studies have investigated using LLMs to apply criteria to study selection in an SS. Therefore, we can extend the \citet{10.1145/3613904.3642216} approach by applying inclusion/exclusion criteria when evaluating papers. Moreover, we can expand further with Retrieval-Augmented Generation (RAG). RAG feeds external knowledge to LLMs as 'context' in the chat. Therefore, we can dynamically update the criteria, leaving only the generic message/reply structure for fine-tuning.

\subsection{Study Quality Assessment}
QA of an SS is also defined as explicit inclusion/exclusion criteria according to  \citet{keele2007guidelines}. Unfortunately, there is no silver bullet for QA checklists, mostly tailored to researchers' investigation topics.
\textit{Classical Approach}
In the classical approach, researchers usually follow the guidelines of  \citet{keele2007guidelines} that provide a list of QA checklists with designs to target for different study designs. Researchers need to choose relevant QA checklists according to the designs of the studies included in the review, like randomized controlled trials, cohort studies, case-control studies, literature reviews, and mapping studies. Each included study is manually evaluated against the relevant QA checklist chosen to point out the methodological strength and related risk of bias.
This step is usually performed by several reviewers who independently check the quality of the study. The studies may be given scores or put into different categories based on the degree of compliance with QA criteria. Higher scores indicate high quality or low risk of bias. The process is quite time-consuming and subjective, and it relies on the interpretation and judgment of humans.

\textit{GAI Approach}
GAI models can be instructed with QA criteria and guidelines so that they can go ahead and perform an independent judgment on collected studies. GAIs will be able to comb huge amounts of papers in a short time, with the ability to respond instantly regarding methodological strengths, biases, and possible concerns. Although GAIs are random in nature (see Limitations) multiple GAIs allows remediating human subjectivity. Moreover, models do not suffer fatigue, thus assuring consistency of application across all studies. Notably, no prior researches have been conducted with respect to the effectiveness of GAIs for QA assessments in SS.


\subsection{Data Extraction}
According to \citet{keele2007guidelines}, data extraction involves collecting and recording relevant information from the selected studies.

\textit{Classical Approach.}
The classical data extraction method involves researchers analyzing selected studies individually to identify and extract the requisite data. Researchers design a standard form or template to document data specifics, i.e., study characteristics, methodological approaches, sample size, results, and other details. Then, this form is pilot-tested on a sub-set of studies to ensure the form captures all data required and is user-friendly. Then, researchers read each study, identify the relevant information, and record this with the extraction form. More than one reviewer should perform data extraction to validate the accuracy and consistency of the data extracted. Hence, researchers must again calculate an IRA to assess biases and misclassification.
This is time-consuming, labor-intensive, and prone to human error. Besides, the process can be overwhelming if the number of studies is large. 

\textit{GAI Approach}
LLM can be fine-tuned or instructed to answer in a specific format and provide or ask to generate the data extraction form. The models can run or create scripts on the fly to automate repetitive activities, such as data extraction from tables, figures, and text.

Moreover, GAI systems can offer immediate feedback and validation and flag possible inconsistencies or missing data in real time. Finally, GAIs allow researchers to process large-scale document collection within a fraction of the time taken by manual methods \cite{esposito2024leveraging}; thus, it is possible to review large-scale literature exhaustively.

\subsection{Data Synthesis }
According to \citet{keele2007guidelines}, data synthesis involves extracting relevant information from the extracted data. Therefore, this last step aims to grasp hidden patterns, research gaps, and current trends in the research landscape. Thus, researchers interpret the synthesized data to draw conclusions and implications based on the context of the SS research questions.

\textit{Classical Approach.}
Classical data synthesis procedures include thematic analysis that extracts common themes and concepts across studies. Or statistical analysis. Depending on the nature of the data, statistical methods may consist of meta-analysis or quantitative synthesis of the findings of studies.

\textit{GAI Approach.}
Finally,  GAI models can easily perform data analysis and synthesis tasks on large-scale studies. The models can process the collected data much faster than using manual methods.   Moreover, GAIs comprehend the semantic context of the collected data and, therefore, can derive the meaning of the text, deducing insightful aspects of the studies. Finally, an advanced GAI model can read, analyze, and output images, tables, and other forms of multimodal data to comprehensively synthesize the original studies' findings.

\section{Roadmap}
\label{sec:roadmap}
\begin{table*}[htpb]
\small
  \centering
  \caption{Metrics and Measurements for Empirical Validation}
    \begin{tabular}{p{4cm}p{5cm}p{5,7cm}}
      \textbf{Step}                  & \textbf{Metrics}                                                       & \textbf{Measurement}                                                                                         \\ \hline
      Identification of Research     & Precision, Recall, F1-score                                            & Compare the relevance of studies retrieved by GAI-generated search strings against manually crafted ones           \\ 
      Publication Bias               & Diversity of sources, Comprehensiveness                                & Evaluate diversity and comprehensiveness of sources retrieved by traditional vs. GAI-based methods             \\ 
      Conducting the Search          & Retrieval precision, Retrieval recall, Search time                      & Assessing the efficiency and effectiveness of GAI-generated search queries compared to manual ones                    \\ 
      Documenting the Search         & Completeness of documentation, Adherence to PRISMA guidelines           & Evaluate completeness and adherence of GAI-generated documentation to PRISMA guidelines                         \\ 
      Study Selection                & Inter-rater agreement (e.g., Cohen's kappa)                            & Measure consistency between human reviewers and GAI models in selecting studies                                 \\ 
      Study Quality Assessment       & Agreement between human reviewers and GAI models (e.g., Cohen's kappa) & Assess agreement in evaluating study quality between human reviewers and GAI models                               \\ 
      Data Extraction                & Precision, Recall, F1-score                                            & Compare accuracy and completeness of data extraction by human reviewers vs. GAI models                            \\ 
      Data Synthesis                 & Coherence of synthesized findings, Insightfulness of interpretations   & Compare coherence and insightfulness of synthesized findings from human reviewers vs. GAI models                
      \\\hline
    \end{tabular}
  \label{tab:metrics-measurements}
\end{table*}
This section sets up the road map for our idea to come to fruition. Our road map consists of two steps: implementation of the framework and empirical validation. Table\ref{tab:metrics-measurements} shows the metrics we plan to use for measuring the effectiveness of the GAI approaches of each SS step.

\textbf{Feasibility}. Our work can contribute to our field by highlighting the current state-of-the-art GAIs-assisted SS tasks and spotting research gaps for future research efforts. Therefore, our vision is based on previous and recent studies assuring its feasibility.

\textbf{Implementation}. We intend to follow up on this vision by implementing a comprehensive suite for autonomous SS. Our implementation will exploit the contributions of the previous studies and new and original contributions as we have presented in Section \ref{sec:design}.

\textbf{Empirical Validation}. We will empirically validate the proof of concept for our framework. Our validation will be based on established guidelines  \cite{DBLP:journals/ese/RunesonH09} and will focus on performing an SLR comparing human results for each SLR step with the GAI ones. More in detail, we foresee a validation protocol for the GAI approach for SS.

Table \ref{tab:metrics-measurements} presents the metrics and the rationale to empirically validate each phases.

\section{Limitations}
\label{sec:limitations}
This section presents the acknowledged limitations to our study and the associated mitigation strategy.

\subsection{Scraping Digital Libraries}
Internet-enabled GAIs can retrieve and scrape websites to retrieve up-to-date information. Unfortunately, publisher agreements can limit scientific publication scraping by limiting its access or denying it altogether. For instance, the \textit{2024-2026 ACM Digital Library Tiered-Band Open Access Model Agreement Terms and Conditions} \footnote{\url{https://libraries.acm.org/binaries/content/assets/libraries/dl/2024-2026-acm-open-blank-agreement-template---consortium_may-2024.docx}} section IV, par. 6, \textbf{allows for computation access} to authorized user. In this context, computational access allows autonomous scraping of scientific publications in the ACM Digital Library with no specific limitations. Nonetheless, GAI is still technically a novel ``issue'' in the publication field. Therefore, publishers can implement measures to limit access specifically to GAI.  We can alleviate this issue by providing a means for the GAI to accept personal or institutional tokens. 
Nonetheless, the scientific community should deeply reflect on the implications of having GAIs scraping scientific knowledge. Unfortunately, the discussion on this topic is too broad to be comprehensively addressed in a section of this vision paper. Nonetheless, we deemed it essential to acknowledge it.

\subsection{Replicability of the findings}
Replicating AI's output is a major issue in AI-based SE research \cite{esposito2024leveraging}. Also, randomness is involved in more straightforward classification tasks. Therefore, the outputs may vary on each use \cite{esposito2024leveraging}. 

Replication packages are essential to reproduce the study analysis \cite{esposito2024validate}. Nonetheless, if randomness is involved, obtaining the same authors' results is challenging.

We can mitigate this limitation when GAI platforms allow for custom setting specific parameters to grant a consistent output, for instance, considering OpenAI API References\footnote{\url{https://platform.openai.com/docs/guides/text-generation/how-should-i-set-the-temperature-parameter}}) two parameters to reduce the randomness of the models and aid in consistency are provided, namely ``\textit{temperature}'' where lower values result in more consistent outputs and ``\textit{seed}'' allowing the control of the initial randomness. 



Moreover, multiple GAI models should be employed to strengthen the reliability. A single model should perform no single step if no prior cross-evaluation occurs. Our empirical validation will also address this issue.

\subsection{Explainability of the GAI models}
LLMs are increasingly employed in essential areas like healthcare, finance, and policy. Therefore, we must ensure that domain experts can effectively collaborate with these models. 

We refer to \citet{arrieta2020explainable} definition of explainable AI:
``\textit{Given an audience, an explainable Artificial Intelligence produces details or reasons to make its functioning clear or easy to understand}.''

Explainability serves as a means to connect human decision-makers with machine learning models, helping to bridge the gap between the two \cite{arrieta2020explainable}.

According to \citet{arrieta2020explainable}, we define our \textit{audience} as the researchers conducting SS. Therefore, any insight and justification made by the explainable model should be easy for them to follow. Transparency and applicability are essential to allow researchers to trust the suggestions and outputs of GAIs used during the entire SLR process \cite{esposito2024leveraging, keele2007guidelines}.

Explainability is hard to achieve due to GAI models' complexity and black-box nature. Thus, while on the one hand, GAI models are most likely to contribute toward increasing the potential for improvements in various aspects of the SLR process, simultaneously, on the other hand, their lack of explainability remains one of the major bottlenecks for the development of trust and collaboration with a community of researchers towards the reasoning behind GAIs output.
It is worth noticing that, due to the volatility of technology, which advances month by month, GAI approaches that today's works may not be tomorrow, and vice versa; they can become obsolete in no time. 
Future studies must build methods and techniques to enhance explainability in GAI models to meet the demand from researchers conducting SS.

\section{Ethical Considerations}\label{sec:ethics}
Our world is fast-changing. Until recently, AI used to categorize things, but now it is \textbf{serving humans imagination}. Therefore, according to \citet{10372461}, SE researchers and practitioners should not ``look away''. 

We need to ensure that GAI models support human researchers and not replace them. They should improve human efficiency by making tedious tasks, such as data extraction and search, quicker and less demanding while allowing researchers more time for critical thinking on higher-level goals and research objectives. Such collaboration would help preserve human oversight and judgments,i.e., the embodiment of ethical considerations, critical thinking, and domain expertise that guide the research process. Transparency and explainability in GAI outputs ensure trust and integrity while conducting SS.

\section{Conclusions}
\label{sec:conlcusions}
In our study, we presented a brief overview of the possible roles of GAI in EBSE to highlight their potential to make several SLR steps more time and resource-efficient alongside a validation protocol.  



\printbibliography

@string{Computing = "Computing" }

@string{Computer = "{IEEE} Computer" }

@string{Springer = "Springer-Verlag" }

@article{arrieta2020explainable,
	title        = {Explainable Artificial Intelligence (XAI): Concepts, taxonomies, opportunities and challenges toward responsible AI},
	author       = {Arrieta, Alejandro Barredo and D{\'\i}az-Rodr{\'\i}guez, Natalia and others},
	year         = 2020,
	journal      = {Information fusion},
	publisher    = {Elsevier},
	volume       = 58,
	pages        = {82--115}
}

@inproceedings{DBLP:conf/esem/SantosMS14,
	title        = {The use of systematic reviews in evidence based software engineering: a systematic mapping study},
	author       = {Ronnie E. S. Santos and Cleyton V. C. de Magalh{\~{a}}es and others},
	year         = 2014,
	booktitle    = {2014 {ACM-IEEE} International Symposium on Empirical Software Engineering and Measurement, {ESEM} '14, Torino, Italy, September 18-19, 2014},
	publisher    = {{ACM}},
	pages        = {53:1--53:4},
	doi          = {10.1145/2652524.2652553},
	editor       = {Maurizio Morisio and Tore Dyb{\aa} and Marco Torchiano},
	bibsource    = {dblp computer science bibliography, https://dblp.org}
}

@inproceedings{DBLP:conf/icse/KitchenhamDJ04,
	title        = {Evidence-Based Software Engineering},
	author       = {Barbara A. Kitchenham and Tore Dyb{\aa} and others},
	year         = 2004,
	booktitle    = {26th International Conference on Software Engineering {(ICSE} 2004), 23-28 May 2004, Edinburgh, United Kingdom},
	publisher    = {{IEEE} Computer Society},
	pages        = {273--281},
	doi          = {10.1109/ICSE.2004.1317449},
	editor       = {Anthony Finkelstein and Jacky Estublier and David S. Rosenblum},
	bibsource    = {dblp computer science bibliography, https://dblp.org}
}

@article{DBLP:journals/ese/RunesonH09,
	title        = {Guidelines for conducting and reporting case study research in software engineering},
	author       = {Per Runeson and Martin H{\"{o}}st},
	year         = 2009,
	journal      = {Empir. Softw. Eng.},
	volume       = 14,
	number       = 2,
	pages        = {131--164}
}

@article{felizardo2020automating,
	title        = {Automating systematic literature review},
	author       = {Felizardo, Katia R and Carver, Jeffrey C},
	year         = 2020,
	journal      = {Contemporary empirical methods in software engineering},
	publisher    = {Springer},
	pages        = {327--355}
}

@article{GAROUSI2019101,
	title        = {Guidelines for including grey literature and conducting multivocal literature reviews in software engineering},
	author       = {Vahid Garousi and Michael Felderer and others},
	year         = 2019,
	journal      = {Information and Software Technology},
	volume       = 106,
	pages        = {101--121},
	doi          = {https://doi.org/10.1016/j.infsof.2018.09.006},
	issn         = {0950-5849}
}

@misc{keele2007guidelines,
	title        = {Guidelines for performing systematic literature reviews in software engineering},
	author       = {Kitchenham, Barbara},
	year         = 2007,
	publisher    = {Technical report, ver. 2.3 ebse technical report. ebse}
}

@article{ghosh2024alpapico,
	title        = {AlpaPICO: Extraction of PICO frames from clinical trial documents using LLMs},
	author       = {Ghosh, Madhusudan and Mukherjee, Shrimon and others},
	year         = 2024,
	journal      = {Methods},
	publisher    = {Elsevier},
	volume       = 226,
	pages        = {78--88}
}

@inproceedings{esposito2024leveraging,
	author = {Esposito, Matteo and Palagiano, Francesco},
title = {Leveraging Large Language Models for Preliminary Security Risk Analysis: A Mission-Critical Case Study},
year = {2024},
isbn = {9798400717017},
publisher = {Association for Computing Machinery},
address = {New York, NY, USA},
doi = {10.1145/3661167.3661226},
booktitle = {Proceedings of the 28th International Conference on Evaluation and Assessment in Software Engineering},
pages = {442–445},
numpages = {4},
location = {Salerno, Italy},
series = {EASE '24}
}

@inproceedings{10.1145/3613904.3642216,
	title        = {EvalLM: Interactive Evaluation of Large Language Model Prompts on User-Defined Criteria},
	author       = {Kim, Tae Soo and Lee, Yoonjoo and others},
	year         = 2024,
	booktitle    = {Proceedings of the CHI Conference on Human Factors in Computing Systems},
	publisher    = {Association for Computing Machinery},
	address      = {New York, USA},
	series       = {CHI '24},
	doi          = {10.1145/3613904.3642216},
	articleno    = 306,
	numpages     = 21
}

@article{10372461,
	title        = {Ethics: Why Software Engineers Can’t Afford to Look Away},
	author       = {Johnson, Brittany and Menzies, Tim},
	year         = 2024,
	journal      = {IEEE Software},
	volume       = 41,
	number       = 1,
	pages        = {142--144},
	doi          = {10.1109/MS.2023.3319768}
}

@article{esposito2024validate,
	title        = {VALIDATE: A deep dive into vulnerability prediction datasets},
	author       = {Esposito, Matteo and Falessi, Davide},
	year         = 2024,
	journal      = {Information and Software Technology},
	publisher    = {Elsevier},
	pages        = 107448
}

@article{PICO,
	title        = {{Evidence-Based Medicine—How to Practice and Teach EBM. D. L. Sackett, W. S. Richardson, W. Rosenberg, and R. B. Haynes. New York: Churchill Livingstone, 1997, 250 pp. Paperback, 24.99. ISBN 0-443-05686-2.}},
	author       = {Henderson, A Ralph},
	year         = 1997,
	month        = 10,
	journal      = {Clinical Chemistry},
	volume       = 43,
	number       = 10,
	pages        = {2014--2014},
	doi          = {10.1093/clinchem/43.10.2014},
	issn         = {0009-9147}
}

@inproceedings{sharma2024generative,
	title        = {Generative Echo Chamber? Effect of LLM-Powered Search Systems on Diverse Information Seeking},
	author       = {Sharma, Nikhil and Liao, Q Vera and others},
	year         = 2024,
	booktitle    = {Proceedings of the CHI Conference on Human Factors in Computing Systems},
	pages        = {1--17}
}

\begin{IEEEbiography}{Matteo Esposito}{\,} is a Researcher at the University of Oulu and has served as the chairman of the ACM Rome Tor Vergata Student Chapter for eight years. His primary research interest lies in Secure Software Engineering, Large Language Models for SSE, and Quantum Software Engineering. He is a PhD Student at the University of Rome Tor Vergata, Italy, where he also earned his MSc and BSc. He is an ACM and IEEE Computer Society member. Contact him at matteo.esposito@oulu.fi.
\end{IEEEbiography}

\begin{IEEEbiography}{Andrea Janes} {\,} is an associate professor at the Free University of Bozen/Bolzano. He was previously a senior lecturer and researcher at the FHV Vorarlberg University of Applied Sciences in Dornbirn, Austria, and a researcher at the Free University of Bozen/Bolzano, Italy.  He holds a Master's degree in Business Informatics from the Vienna University of Technology and a PhD in Computer Science (with honors) from the University of Klagenfurt, Austria. He obtained the habilitation in Computer Science and Information processing systems. He is particularly interested in Lean and Agile approaches to software engineering, value-based software engineering, empirical software engineering, software testing, and technology transfer. Contact him at andrea.janes@unibz.it.
\end{IEEEbiography}

\begin{IEEEbiography}{Davide Taibi} {\,} is a Full Professor at the University of Oulu, where he heads the M3S Cloud research group. His research is mainly focused on Cloud and Edge Software Architecture. In particular, my research group is working on migration to cloud-native technologies; Good practices, patterns, and anti-patterns for cloud-native technologies, including edge computing, serverless, and microservices; Architectural Degradation and Architectural Technical Debt - Quality of Open Source Software. Prediction of Open Source Adoption Risk, Open Source Abandonment Prediction. He has been a member of the International Software Engineering Network (ISERN) since 2018. Formerly, he worked at the Free University of Bolzano, Technical University of Kaiserslautern, Germany, Fraunhofer IESE - Kaiserslautern, Germany, and Università degli Studi dell’Insubria, Italy. In 2011 he was one of the co-founders of Opensoftengineering s.r.l., a spin-off company of the Università degli Studi dell’Insubria. Contact him at davide.taibi@oulu.fi.
\end{IEEEbiography}

\begin{IEEEbiography}{Valentina Lenarduzzi} {\,} is an Associate Professor (Tenure Track) at the University of Oulu, Finland, who specializes in data analysis in software engineering, with a focus on software quality, maintenance, and evolution. She holds a Ph.D. in Computer Science from the University of Insubria (Italy), where she researched effort estimation and data analysis in software engineering. Valentina has also held postdoctoral positions at LUT University (Finland), Tampere University (Finland), and the Free University of Bozen-Bolzano (Italy). She has contributed as a visiting researcher at institutions such as the Technical University of Kaiserslautern and the Fraunhofer Institute for Experimental Software Engineering IESE (Germany). Contact her at valentina.lenarduzzi@oulu.fi.
\end{IEEEbiography}
\end{document}